\documentclass[letterpaper,11pt]{article}
\usepackage[T1]{fontenc}
\usepackage[utf8]{inputenc}
\usepackage[margin=1in]{geometry}
\usepackage{amssymb}
\usepackage{amsmath}
\usepackage[english]{babel}
\usepackage{algorithm}
\usepackage[noend]{algpseudocode}
\usepackage{todonotes}
\usepackage{xspace}

\usepackage{amsmath,amssymb,amsthm}
\usepackage{mathtools}
\usepackage{thmtools}
\usepackage{amsfonts}
\usepackage{graphicx}
\usepackage{sidecap}
\usepackage{lineno}
\usepackage[noadjust]{cite}

\usepackage{hyperref}
\hypersetup{
    colorlinks=false,
    linkcolor=blue,
    urlcolor=cyan,
}
\usepackage{cleveref}

\declaretheorem[numberwithin=section]{theorem}

\declaretheorem[numberwithin=section]{lemma}
\declaretheorem[numberwithin=section]{definition}

\newcommand{\E}[1]{{\mathbb E \left[ #1 \right]}}
\newcommand{\Prob}[1]{{\mathbb P \left[ #1 \right]}}

\newcommand{\Oh}{\mathcal{O}}

\DeclarePairedDelimiter\abs{|}{|} 
\DeclarePairedDelimiter\pa{(}{)} 
\newcommand{\query}{\textsc{Query}}
\newcommand{\polylog}{\mathrm{polylog\,}}
\DeclareMathOperator{\dep}{depth}
\DeclarePairedDelimiter\ceil{\lceil}{\rceil}

\algnewcommand{\LineComment}[1]{\State \(\triangleright\) #1}

\usepackage{amsthm}
\usepackage{authblk}

\title{Better Bounds For Graph Reconstruction\footnote{Partially supported by the Polish National Science Centre grant number 2023/51/B/ST6/01505.}}
\author[1]{Antoni Buraczewski}
\author[1]{Pawe\l{} Gawrychowski}
\affil[1]{Institute of Computer Science, University of Wrocław, Poland}
\date{}
\begin{document}

\maketitle

\begin{abstract}
The goal in graph reconstruction problems is to learn an initially unknown graph using only a few queries
of a specific type. Possibly the most natural query type is retrieving the distance between any two vertices
of a connected, undirected, and unweighted graph. For a bounded-degree graph on $n$ vertices,
Kannan, Mathieu, and Zhou [TALG 2018] showed a randomized algorithm that uses only $\tilde\Oh(n^{3/2})$
such queries. On the lower bound side, it is only known that any randomized algorithm needs
$\Omega(n\log n)$ queries.
We make a significant step forward to bridging these bounds by designing a randomized algorithm
that uses $n^{4/3+o(1)}$ queries. This requires extending the original approach based on Voronoi
diagrams into a recursive procedure that partitions a cell into smaller cells.
While such a recursive approach has been used in the related graph verification problem, in the more general
graph reconstruction problem we need to operate without knowing the graph, which requires a new
(and more technical) approach.
\end{abstract}

\section{Introduction}

Consider a large unknown graph on $n$ vertices that can only be accessed by querying for the distance between any two vertices.
We want to reconstruct the whole graph with only a few such queries. This naturally captures the setting where
the queries are slow or expensive. We stress that we want to minimise the number of queries, and do not concern ourselves with
the time necessary to actually reconstruct the graph based on the answers (nor the time to figure out which query should be
asked next). However, ideally it should be polynomial in the size of the graph.

There are at least two interesting scenarios in which we might be interested in such a task. First, we have the network reconstruction
problem~\cite{DBLP:journals/jacm/AchlioptasCKM09,DBLP:journals/jsac/BeerliovaEEHHMR06,DBLP:journals/tcs/DallAstaABVV06,5072135},
in which we want to recover a hidden network topology using a tool such as \texttt{traceroute}.
Second, we have the evolutionary tree recovery problem~\cite{Reyzin2007LongestPath,DBLP:conf/soda/KingZZ03,Hein1989AnOA,waterman1977additive},
in which we want to recover a hidden evolutionary tree by computing similarities between some species.

Reyzin and Srivastava~\cite{DBLP:conf/alt/ReyzinS07} provided an elementary information-theoretical argument establishing that
reconstructing a general (connected, undirected, and unweighted) graph on $n$ vertices requires $\Omega(n^{2})$ queries.
The lower bound already holds for a constant-depth tree. However, the difficult instance contains a root of degree $\Omega(n)$.
This motivated Kannan, Mathieu, and Zhou~\cite{DBLP:journals/talg/KannanMZ18} to consider the more restricted but still well-justified
from the point of view of the above applications case of bounded-degree graphs. Under such an assumption, they showed
that $\tilde\Oh(n^{3/2})$\footnote{In this paper, $\tilde\Oh(f(n))$ denotes $\Oh(f(n)\polylog n)$.} queries are always sufficient in expectation to
recover the graph (that is, the algorithm always correctly recovers the graph, and the number of queries is a random variable
with the expected value upper bounded by $\tilde\Oh(n^{3/2})$).
However, they were only able to show that $\Omega(n\log n/\log\log n)$ queries are necessary for such graphs.
For random regular graphs (with small degree), Mathieu and Zhou~\cite{DBLP:journals/rsa/MathieuZ23} showed that
$\tilde\Oh(n)$ queries are enough (and, in fact, the same simple procedure recovers an arbitrary bounded-degree graph with
$\tilde\Oh(n^{5/3})$ queries).
Krivelevich and Zhukovskii~\cite{krivelevich_et_al:LIPIcs.ESA.2025.30} considered the same question in the more
standard binomial random graph model $G(n,p)$ and obtained both lower and upper bounds parameterized by the density.
Kannan, Mathieu, and Zhou~\cite{DBLP:journals/talg/KannanMZ18} considered the reconstruction problem for chordal and outerplanar
graphs, showing that in both cases $\tilde \Oh(n)$ queries are enough (see also~\cite{DBLP:journals/tcs/RongLYW21} for better bounds for chordal graphs).
Bastide and Groenland~\cite{DBLP:journals/rsa/BastideG25} considered the reconstruction problem for trees, chordal graphs, and graphs with bounded treelength,
obtaining tight results. Interestingly, they were able to establish that any randomized algorithm requires $\Omega(n\log n)$ queries
to reconstruct a bounded-degree tree on $n$ nodes, which remains the highest known lower bound for this problem, leaving
a large gap between the lower bound of $\Omega(n\log n)$ and the upper bound of $\tilde\Oh(n^{\frac 32})$. Eppstein, Goodrich, and Liu~\cite{DBLP:conf/esa/EppsteinGL25} considered reconstructing graphs with small BFS width, and Kaudan and Nayyeri~\cite{DBLP:conf/swat/KaudanN26} considered reconstructing graphs with bounded cutwidth.

Kannan, Mathieu, and Zhou~\cite{DBLP:journals/talg/KannanMZ18} showed that
natural greedy algorithms are enough to recover the graph with $n^{1+o(1)}$ queries that
return not only the distance but the actual shortest path, and to verify a given graph with $n^{1+o(1)}$ distance queries.
This suggests that perhaps $n^{1+o(1)}$ queries are enough to recover an arbitrary bounded-degree graph with distance queries,
but decreasing the upper bound of $\tilde\Oh(n^{3/2})$ remains an important open problem in this area.

In an attempt to understand the limitations of the known algorithms for reconstructing a graph with distance queries,
Abrahamsen, Bodwin, Rotenberg, and Stöckel~\cite{DBLP:conf/stacs/AbrahamsenBRS16} introduced the weaker betweenness
queries, which given three vertices $x, y, z$ return whether $y$ lies on a shortest path between $x$ and $z$. Surprisingly,
they were able to show a randomized algorithm that uses only $\tilde\Oh(n^{3/2})$ queries. This suggests that improving the
bound on the number of distance queries should exploit the actual distances.

\paragraph{Our results} We design an algorithm that recovers an arbitrary bounded-degree graph on $n$ vertices with only
$n^{4/3+o(1)}$ distance queries in expectation. More formally, we show the following theorem.

\begin{theorem}
\label{thm:main}
There is a randomized algorithm that, given access to a connected $n$-vertex graph in which every vertex has degree at most $\Delta$
through queries that return the distance between any two vertices, recovers the graph using
$n^{4/3} \cdot \pa{\Delta\log n}^{\Oh\pa{\log\log n}}$
queries in expectation.
 \end{theorem}

\noindent In particular, for $\Delta=\polylog n$ this is $n^{4/3+o(1)}$.

We stress that we will measure the complexity of all algorithms only in the number of queries. However, it is actually not difficult to verify
that our algorithm can be implemented efficiently, that is, in polynomial time. Further, when presenting some of the procedures
we will only show that the claims on their correctness and running time hold with constant probability. However, in every
recursive call the procedures will
be repeated until they achieve the desired guarantees on the sizes of the cells, so in fact the whole algorithm is Las Vegas.

\paragraph{Related work} There is a vast body of work concerned with reconstructing an $n$-vertex graph with queries of different types.
Perhaps the simplest are independent set queries~\cite{DBLP:journals/dam/GrebinskiK98}, where we can query whether, for a set of
vertices $S$, the subgraph induced by $S$ contains at least one edge. It is known that in this model $\Oh(\log n)$ queries
per edge suffice~\cite{DBLP:journals/jcss/AngluinC08}. The more powerful additive queries return the number of edges
in the subgraph induced by $S$, allowing one to achieve an asymptotically optimal bound for dense~\cite{DBLP:journals/algorithmica/GrebinskiK00} and
sparse~\cite{DBLP:conf/stoc/ChoiK08} graphs, and also some restricted classes of graphs~\cite{DBLP:journals/siamcomp/AlonBKRS04}.
Queries that return the number of connected components~\cite{DBLP:conf/colt/BlackMSX25} or an adversarially chosen maximal
independent set~\cite{DBLP:conf/innovations/0001OT25} have also been considered.
Another example is cut queries that, for two disjoint sets of vertices $S$ and $T$, return the number of edges with one
endpoint in $S$ and the other in $T$. However, such queries are usually considered in the context of connectivity
(for example, computing the global min cut, or the $s$-$t$ cut)~\cite{DBLP:conf/innovations/RubinsteinSW18,DBLP:conf/focs/ApersEGLMN22,DBLP:conf/soda/JiangNS26,DBLP:conf/soda/Kenneth-Mordoch26,DBLP:conf/esa/Kenneth-Mordoch25,DBLP:conf/soda/LeeSZ21}.
A separate line of work concerns edge queries, see~\cite{LovaszYoung91} for a thorough introduction.

\section{Preliminaries}

\paragraph{Graphs} We denote the (connected) input graph by $G=(V,E)$, where $n=|V|$.
We assume that the degree of each vertex is upper bounded by $\Delta$. Throughout most of the paper to avoid clutter
we will assume that the graph has bounded degree, that is, $\Delta=\Oh(1)$, thus $\tilde\Oh\pa{\cdot}$ will hide factors
polylogarithmic in $n$ under such an assumption. However, we will analyse the exact dependency on $\Delta$ in our final algorithm.

\paragraph{Distances} For two vertices $u,v\in V$ we write $\delta(u, v)$ to denote the distance between $u$ and $v$ in the whole $G$.
For a vertex $u\in V$ and a set of vertices $S\subseteq V$ we write $\delta(u, S)$ to denote the distance between $u$ and $S$ in the whole $G$,
defined as the minimum of $\delta(u,v)$ over all $v\in S$.

\paragraph{Neighborhoods} We use the following notation for the neighborhood and $2$-neighborhoods of a vertex $v\in V$, respectively:
$N (v) = \{u \in V : \delta (u, v) \leq 1\}$ and $N_2(v) = \{u \in V : \delta (u, v) \leq 2\}$.

\paragraph{Queries} We assume that $n$ and $\Delta$ are both known to the algorithm (as observed in some of the prior work, the latter
assumption is not strictly necessary, as if the number of queries significantly exceeds our estimate then we can restart the algorithm
while doubling the value of $\Delta$). No other information about the graph is given, but we can query for the distance between any
two vertices $u,v\in V$. Such a query is denoted by $\query(u, v)$. We can also query for the distances between any vertex $u\in V$
and all vertices in $S\subseteq V$, which is denoted by $\query(u, S)$. We stress that the cost of such a query is $|S|$.
Similarly, we can query for the distances between any vertex in $S\subseteq V$ and any vertex in $T\subseteq V$, which is denoted by
$\query(S, T)$. The cost of such a query is $|S| \cdot |T|$.

\paragraph{Clusters} For a set $S \subseteq V$ we define the cluster of a vertex $w\in V$ with respect to a set of vertices $A\subseteq V$ as
$C_A (w) = \{v \in S : \delta (w, v) < \delta(v, A)\}$. We will often talk about the vertices with large clusters,
defined as $W_A = \{v \in V : \abs{C_A(v)}\geq \frac{6\abs{S}}t\}$ (where $t$ is a parameter).

\paragraph{Sampling} For a parameter $t$ and a set of vertices $S\subseteq V$ we write
$\textsc{Sample}(S, t)$ to denote sampling each vertex from $S$ with probability $\frac t{\abs{S}}$. Sometimes we will need slightly different distributions, namely sampling a uniformly random set (or multi-set) of exactly $t$ vertices.

\paragraph{Probabilistic tools} As usual, for a random variable $X$ we write $\E X$ to denote its expected value.
We need the following standard multiplicative form of the Chernoff bound: for any random variable $X$ equal to the sum of independent $0$-$1$ random variables, we have:
\[\Prob{X \leq (1-\delta)\E{X}} \leq e^{-\delta^2\E X /2} \] 
for any $\delta\in [0,1]$.

\section{Technical Overview}

The trivial quadratic algorithm uses the observation that two vertices are connected by an edge if and only if the distance between them is equal to $1$. We refer to this approach as an exhaustive search. We perform it only on sufficiently small subsets.

Rather than focusing on the whole graph, we try to find the edges induced by a particular subset $S$. Extending the work done by Kannan, Mathieu, and Zhou~\cite{DBLP:journals/talg/KannanMZ18}, we are going to design a recursive algorithm that aims to reduce searching for the edges induced by a subgraph to searching in a few smaller induced subgraphs. In contrast to~\cite{DBLP:journals/talg/KannanMZ18} where the graph is partitioned only once, our approach allows for arbitrary induced subgraphs, which enables the use of recurrence. In fact, the procedures shown in this section were initially presented for $S = V$. However, the analysis for a general $S \neq V$ remains the same so we present it in this more general setting. While these procedures are not sufficient to design a recursive algorithm, they are necessary to understand our approach. The proof of the generalized (to $S \neq V$) procedure is only a minor modification of the proof of \textsc{Theorem~3.1} from~\cite{DBLP:conf/spaa/ThorupZ01}, already discussed in the analysis of \textsc{Algorithm~4} from~\cite{DBLP:journals/talg/KannanMZ18}.

Covering the graph induced by $S$ with smaller subgraphs relies on the concept of Voronoi cell decomposition. The polynomial time algorithm which ensures that smaller subgraphs have bounded size was designed by Thorup and Zwick~\cite{DBLP:conf/spaa/ThorupZ01} and extended to the distance oracle model by Kannan, Mathieu, and Zhou~\cite{DBLP:journals/talg/KannanMZ18}.

\subsection{Voronoi cells}

Consider a subset $S \subseteq V$ and a subset $A \subseteq V$ called the set of centers. Observe that we do not require $A \subseteq S$. In fact, it is crucial for the algorithm that some centers are outside $S$. Similarly to the Voronoi diagram on the plane, we try to cover the vertices of $S$ by regions. Each region is represented by a vertex $a \in A$ and consists of vertices for which $a$ is roughly the closest vertex in $A$. This should be done in a way that covers all edges in the subgraph induced by $S$ to allow us to recurse on each cell. Formally, the following two characterizations of Voronoi cells will be used. The vertex $v \in S$ belongs to the cell centered at $c \in A \subseteq V$ if either of the two equivalent
conditions is satisfied:
\begin{itemize}
	\item $v \in N(c)$ or $v \in C_A(u)$ for some $u \in N_2(c)$,
	\item $\delta(v, c) \leq \delta(v, A) + 1$.
\end{itemize}
The second condition is more intuitive: it says that the cell of $c$ contains vertices for which $c$ is almost the closest vertex from $A$. The equivalence follows from considering a vertex $u \in N_2(c)$ which is on the shortest path from $v$ to $c$ at distance $2$ from $c$. Although the latter characterization is more similar to the concept of a Voronoi diagram, we will mostly use the former. It gives us control over the size of each cell by controlling the sizes of $C_A$. The definition implies that, in particular, a vertex may belong to more than one cell. Henceforth, when we refer to partitioning or dividing a subset we actually mean covering all edges in the subgraph induced by a subset with smaller overlapping Voronoi cells. Note that if we want to cover each edge of $S$, the intersections of some cells have to be nonempty.

From the second condition, it can easily be seen that for every edge $e=(u,v)$ with $u,v \in S$ there exists a cell that contains both $u$ and $v$, without any assumptions on $S$. Specifically, if $\delta(u, A) \leq \delta(v, A)$ and $c_u \in A$ is closest to $u$ then $u, v$ belong to the cell of $c_u$
(which is why we have the $+1$ in the second condition).
Therefore, to obtain a subquadratic algorithm, it suffices to find the set of centers $A$ that divides $S$ into Voronoi cells of ``small'' sizes and for each of them perform the exhaustive search. We can define ``small'' formally. Let $\abs A = \tilde\Oh\pa t$. Since each vertex in $S$ belongs to at least one cell, the size of the largest cell must be at least $\tilde\Omega\pa{\frac {\abs S}t}$. Hence, we can hope for cells of sizes $\tilde\Oh\pa{\frac {\abs S}t}$.
We will show in the next subsection that this can be achieved using a randomized algorithm.

If the set of centers $A \subseteq V$ is known, the pairwise distances between $A$ and $S$ are all that is needed to find the cells. This follows directly from the second condition in the definition of cells. 
In the distance oracle model, we can simply learn all the distances between $A$ and $S$. Hence, the Voronoi cells for $A$ can be computed using $\abs{A} \cdot \abs{S}$ distance queries. However, such an algorithm is not actually very useful for our purposes, and we will apply another approach
that uses the first definition of Voronoi cells.
With an additional assumption that $N_2(A)$ is known, we can construct an algorithm which uses the notion of clusters,
see Algorithm~\ref{alg:cells}. It in particular illustrates the fact that if the clusters are small, the whole cell will be small.


\begin{algorithm}
\caption{Dividing a set $S$ using Voronoi cells centered at $A$.}
\label{alg:cells}
\begin{algorithmic}[1]
	\Function{Voronoi-Cells}{$S, A$}
	\LineComment{$N_2(a)$ needs to be known for every $a\in A$.}
	\LineComment{We don't necessarily have $A \subseteq S$.}
		\State $\mathcal{D} \leftarrow \emptyset$
		\State $\query(A, S)$
		\For{$a \in A$}
			\State $\query(N_2(a), S)$
			\For{$b \in N_2(a)$}
				\State $C(b) \leftarrow \{v \in S : \delta(b, v) < \delta(v, A)\}$
			\EndFor
			\State $D_a \leftarrow \bigcup \{C(b) : b \in N_2(a)\} \cup \pa{N(a) \cap S}$
			\State $\mathcal{D} \leftarrow \mathcal{D} \cup \{D_a\}$
		\EndFor
		\State \Return $\mathcal{D}$
	\EndFunction
\end{algorithmic}
\end{algorithm}

\subsection{Finding small Voronoi cells of a known graph from~\cite{DBLP:conf/spaa/ThorupZ01}}

Let us first focus on a setting where all the distances between $S$ and $V$ are known. This will have to be later adjusted to the distance oracle model in a way that allows us to use less than $\abs S \cdot \abs V$ queries. A procedure (Algorithm~\ref{alg:thorup}) to find a partition of a subset $S$ with small cell sizes was given by Thorup and Zwick~\cite{DBLP:conf/spaa/ThorupZ01}. It keeps track of the set $W_A = \{v \in V : \abs{C_A(v)}\geq \frac{6\abs S}t\}$\footnote{The constant $6$ was chosen arbitrarily. It is only necessary that it is strictly bigger than $2$.}
while constructing the set $A$ of size $\tilde\Oh(t)$. The subscript $A$  in $W_A$ is omitted if the set $A$ is clear from the context. The reason for considering $W$ lies in the definition of the Voronoi cell centered at $a \in A$ as $\bigcup_{u\in N_2(a)}{C_A(u)} \cup \pa{N(a) \cap S}$. Hence, if $N_2(a) \cap W = \emptyset$, the size of the cell at $a$ is at most
\begin{align*}
\abs{N(a)} + \sum_{u\in N_2(a)}{\abs{C_A(u)}} &\leq \abs{N(a)} + \abs{N_2(a)}\cdot \frac {6\abs S}t \\
&\leq \Delta + 1 + \pa{\Delta^2+1}\cdot\frac{6\abs S}{t} \\
&= \tilde\Oh\pa*{\frac {\abs S}t}
\end{align*}
which is the best possible bound, ignoring the $\tilde\Oh\pa{1}$ factors.
Therefore, we want to find a set $A$ for which $N_2(A) \cap W = \emptyset$.

It was shown already by Thorup and Zwick~\cite{DBLP:conf/spaa/ThorupZ01} that when we sample each vertex from $W$ with probability $\frac t{\abs{W}}$ and add them to $A$, the new size of $W$ after recomputation will be, with probability at least $\frac 12$, at least $2$ times smaller than the previous one. In our algorithm (and in the presentation of the previous works) we use a slightly different distribution, that is, sampling exactly $t$ vertices from $W$. This is a slight change, which also guarantees halving of $W$ with probability at least $\frac 12$. We discuss this issue in~\Cref{sec:distributions}.

To ensure that $W$ becomes empty, it suffices to perform only logarithmically many iterations in expectation. Then, if after some iteration $W = \emptyset$, we have $N_2(A) \cap W = N_2(A) \cap \emptyset = \emptyset$ and we found a partition into small Voronoi cells. In each iteration we add $t$ elements to $A$, thus the expected total number of cells is $\abs{A} = \Oh\pa{\log n} \cdot t = \tilde\Oh\pa{t}$. As desired, the maximal cell size is approximately the same as optimal.

\begin{algorithm}
\caption{Selecting centers as in~\cite{DBLP:conf/spaa/ThorupZ01}.}
\label{alg:thorup}
\begin{algorithmic}[1]
	\Function{Centers}{$S, t$}
	\State $A \leftarrow \emptyset, W \leftarrow V$
	\While{$W \neq \emptyset$}
		\State $A' \leftarrow$ random $t$-element subset of $W$
		\LineComment{or $A' = W$ if $\abs{W} < t$}
		\State $A \leftarrow A \cup A'$
		\For{$w \in W$}
			\State $C(w) \leftarrow \{v \in S : \delta(w, v) < \delta(v, A)\}$
		\EndFor
		\State $W \leftarrow \{w \in W : \abs{C(w)} \geq \frac{6\abs S}t\}$
	\EndWhile
	\State \Return $A$
	\EndFunction
\end{algorithmic}
\end{algorithm}

\subsection{Finding centers by estimation (for unknown graphs) from \texorpdfstring{\cite{DBLP:journals/talg/KannanMZ18}}{[KMZ18]}}

Kannan, Mathieu, and Zhou~\cite{DBLP:journals/talg/KannanMZ18} showed that the algorithm described in the previous subsection can be modified to use a subquadratic number of queries. The crucial change is related to the calculation of the size of the cluster of a given vertex and the set $W$.
Maintaining the set $W$ without knowing all the distances between $A$ and $S$ looks hard. Hence, instead of finding the exact size of a cluster, we only estimate its size by sampling sufficiently many vertices. Denote this estimate by $\tilde C(v)$. Based on that, we can keep track of the set $\tilde W$, defined as the set of vertices with large \emph{estimation} of a cluster size. By sampling a sufficiently large random subset of the vertices, we can claim that with high probability $\tilde C(v) > \frac {5\abs{S}}t$ implies $\abs{C(v)} > \frac {4\abs{S}}t$, and similarly that $\tilde C(v) < \frac {5\abs{S}}t$ implies $\abs{C(v)} < \frac {6\abs{S}}t$. That is, with high probability the estimation is reasonable. Hence, if we discard a vertex from $\tilde W$, it is very likely that it has a small cluster size and that it is also outside the current set $W$ (which is never maintained by the algorithm). Therefore, an analysis similar to that of~\Cref{alg:thorup} implies that $\tilde W$ halves with constant probability.

Note that we distinguish between $W$ and $\tilde W$. The former is defined precisely (deterministically) as an abstract object, while the latter is its estimation, a random variable defined by running the algorithm. By the above observation about the correlation between the values of $\tilde C(v)$ and $\abs{C(v)}$, we can deduce that w.h.p. $W \subseteq \tilde W$. Hence, although we do not directly use the set $W$, we can still control its size. In particular, the algorithm finishes when $\tilde W = \emptyset$, but w.h.p. this implies that also $W$ is empty.
Correctness can be established by similar arguments, leading to the following:

\begin{lemma}[Lemma~2.3,~\cite{DBLP:journals/talg/KannanMZ18}]
	\label{lem:alg_basic_proof}
	With probability at least $\frac 14$, the following events related to the running of \Cref{alg:basic} (\textnormal{\textsc{Estimated-Centers}}) happen simultaneously:
	\begin{itemize}
		\item after every iteration the condition $W_A \subseteq \tilde W$ holds,
		\item the algorithm terminates in $12\log n$ iterations using $\tilde\Oh\pa{nt}$ queries,
		\item and partitions $S$ into cells of size $\tilde\Oh\pa{\frac{\abs{S}}t}$ by choosing a set of centers $A \subseteq V$ with $\abs{A} = \tilde\Oh\pa{t}$.
			\end{itemize}
\end{lemma}

\begin{algorithm}
\caption{Selecting centers by estimation}
\label{alg:basic}
\begin{algorithmic}[1]
	\Function{Estimated-Centers}{$S, t$}
	\State $A \leftarrow \emptyset, \tilde W \leftarrow V$
	\State $T \leftarrow K \cdot t \cdot \log n$ \Comment{$K$ is a large enough constant}
	\While{$\tilde W \neq \emptyset$}
		\State $A' \leftarrow$ random $t$-element subset of $\tilde W$
		\LineComment{or $A' = \tilde W$ if $\abs{\tilde W} < t$}
		\State \query$(A', S)$
		\State $A \leftarrow A \cup A'$
		\For{$w \in \tilde W$}
			\State $X \leftarrow$ random multi-subset of $S$ of size $T$
			\State \query$(X, w)$
			\State $\tilde C(w) \leftarrow \abs{\{v \in X : \delta(w, v) < \delta(v, A)\}} \cdot \frac {\abs{S}}T$
		\EndFor
		\State $\tilde W \leftarrow \{w \in \tilde W : \tilde C(w) \geq \frac{5\abs S}t\}$
	\EndWhile
	\State \Return $A$
	\EndFunction
\end{algorithmic}
\end{algorithm}

The intuition behind partitioning a subset to obtain a better algorithm is as follows.
If we choose $t = s$ and $S = V$ then we can partition $V$ into $\tilde \Oh\pa s$ cells of total size $\tilde\Oh\pa n$ using $\tilde\Oh\pa{ns}$ queries. The largest of them will have a size $\tilde\Oh\pa{\frac ns}$. Each edge will be contained in at least one cell, so we can just perform the exhaustive search for each cell. The algorithm uses $\tilde\Oh\pa{n\cdot \frac ns}$ queries (the sum of each size times the size of the largest cell). We can balance the two components by picking $s = \Theta\pa{n^{1/2}}$ to get the complexity of $\tilde\Oh\pa{n^{3/2}}$.

\subsection{Challenges in extending prior work}

The main result of this paper is showing how to use recursion to partition each cell into smaller ones. That is, there is a set of active subsets of vertices with the property that each edge is contained in at least one active subset. We want to take one active subset, partition it using Voronoi cells and deactivate it in favor of the new cells. Let us discuss the challenges that have to be faced by this approach. First, we argue why we cannot simply call the procedure from the previous subsection recursively. The basic reason is that the problem for a subset is not equivalent to the problem on a whole graph because the shortest paths are calculated in the whole graph. Thus, we cannot ignore the rest of the graph. This was formalized by Bastide~\cite[Theorem~2.7]{bastide:tel-05262763}, who constructed a family of bounded-degree graphs $\{\mathcal G_i\}$ and their subsets $S_i \subseteq V\pa{\mathcal G_i}$ such that any randomized algorithm reconstructing the subgraph induced by $S_i$ using distance queries only between vertices from $S_i$ has to use at least $\tilde\Omega\pa{\abs{S_i}^2}$ queries in expectation. Moreover, the set $S_i$ is natural: it could be a Voronoi cell found by~\Cref{alg:basic} (from~\cite{DBLP:journals/talg/KannanMZ18}) called on the set of all vertices.

Another simple idea for using recursion would be to take the subset of centers $A$ partitioning $S$ to be a subset $A \subseteq S$. However, this approach also turns out to be impossible, which can be justified in two ways:
\begin{itemize}
	\item We impose no structural assumptions on the cells. It could happen that for some set $S \subseteq V$ and parameter $t$ there is no way to choose a subset $A \subseteq S$ of size $\abs{A} = \tilde\Oh\pa{t}$ such that all cells centered at $A$ have sizes $\tilde\Oh\pa{\frac{\abs{S}}t}$. This can be formally proven by constructing $S$ such that there is a vertex $v \in S$ for which there exist many vertices in $S$ which are pairwise far away in the subgraph induced by $S$ but all of them are close to $v$ in the whole graph $G$. For the constructed graph, most of these vertices (together with some other vertices close to them) lie in the same cell as $v$, regardless of the choice of $A \subseteq S$. This way, the cell containing $v$ becomes too large. We provide more details on such a construction in~\Cref{sec:counterexample}.
		
	\item Even if there is a way to prevent the above issue (by some manipulation with the centers in the previous level of recursion), the usage of Voronoi cells is limited. The main method to bound the size of a cell centered at $a \in A$ relies on bounding the size of clusters of $N_2(a)$. However, these neighbors can be outside of $S$, even if $a \in S$. Moreover, the algorithm for partitioning into cells takes vertices to $A$ from the whole subset of vertices $W$ for which it controls the cluster sizes. Thus, we have to control the sizes of clusters of vertices outside of $A$.
\end{itemize}

Thus, vertices outside of $S$ have to be considered for being the centers of cells, and it is not clear how to restrict the set of candidates for such centers.
Therefore, we assume only that $A \subseteq V$. It is worth mentioning that there is a simpler version of the problem that uses recursion, namely the verification problem: we are given a (bounded degree) graph $G$ and a distance oracle to a graph $\tilde G$. We have to check if $G = \tilde G$. This was solved by Kannan, Mathieu and Zhou in~\cite{DBLP:journals/talg/KannanMZ18} using $n^{1+o(1)}$ queries in expectation. Their algorithm recursively found the partition of $G$ into Voronoi cells and compared whether $\tilde G$ partitions in the same way. This is similar to the reconstruction problem but is significantly easier: the algorithm for verification searches for a partition into cells ``offline'', without using any queries. In the reconstruction problem, we have to find a partition relying only on information received from the queries.

As mentioned earlier, Abrahamsen, Bodwin, Rotenberg, and Stöckel~\cite{DBLP:conf/stacs/AbrahamsenBRS16} introduced the weaker betweenness oracle in order
to study the limitations of the distance oracles. In particular, they observed that a limitation of the previous algorithm (as their algorithm using the betweenness queries
follows the blueprint of Kannan, Mathieu and Zhou in~\cite{DBLP:journals/talg/KannanMZ18}) is that, whenever it compares the distance between $u$ and two vertices
$v$ and $v'$, it actually has access to the distance between $u$ and every other vertex. This seems problematic when trying to extend the algorithm to be recursive.
In particular, in the betweenness oracle model, retrieving the distance appears to require retrieving the whole shortest path, and in the recursive approach the shortest
paths are not necessarily contained in the current cell, and in fact might be much longer than the cell.

\subsection{This paper}

We develop new techniques for effectively partitioning a subset $S\subseteq V$ into Voronoi cells. Our main new contribution is a method that reduces the number of candidates for
the cell centers.
We proceed in two phases. The first uses a modification of the earlier approach: while the earlier approach estimates cluster sizes for all vertices, we only do so for a random subset of vertices. The subset needs
to be not too small (as otherwise we might not collect enough vertices to subpartition the current cells) and not too large (to avoid issuing too many queries).
The set of centers generated this way is not yet guaranteed to be good, but it turns out that the number of candidates for the remaining centers can be significantly
reduced to the vertices near (in some sense) $S$. This enables us to efficiently identify the remaining candidates and perform the rest of the search using a small number
of queries.
To make such an approach work, it turns out that it is necessary to impose an additional restriction on the Voronoi cells: we need to bound not only the size of each
cell but also the number of cells containing each vertex. The latter was not necessary in prior work, and requires a new tool.
We design such a tool that, informally speaking, fixes the set of centers by recursively ``spreading'' it randomly to the neighbors.
While such a tool can be easily combined with the previous algorithm for finding Voronoi cells, combining it with our recursive approach turns out to be non-trivial.

Prior to discussing the details of how to partition a subset $S$ into cells, we present the high-level structure of the whole algorithm.
The desired query complexity is $\tilde\Oh\pa{ns}$, where $s$ is a threshold fixed throughout the algorithm.
Thus, at each recursion depth we are allowed to use $\tilde\Oh\pa{ns}$ queries in total, so to partition $S$ we can use $\tilde\Oh\pa{\abs{S}s}$ queries.
It turns out that our algorithm can use that many queries to decrease the size of a cell from $\abs{S}$ to $\tilde\Oh\pa{\abs{S}^{\frac 23} \cdot \pa{\frac{n}{s^2}}^{\frac 13}}$ (or, in the non-recursive approach, to $\tilde\Oh\pa*{\sqrt{\abs{S} \cdot \frac{n}{s^2}}}$).
Hence, our final cell sizes will be $\tilde\Oh\pa{\frac{n}{s^2}}$ and the recursion depth is $\Oh\pa{\log\log n}$. The bound on the depth allows us to not worry
about the $\tilde\Oh\pa{1}$ factors, since they are at most $\pa{\log n}^{\Oh\pa{\log\log n}} \leq n^{o(1)}$. The final exhaustive search over all cells will take $\Oh\pa{n \cdot \frac n{s^2}}$
(ignoring the subpolynomial factors), so after comparing it to $\Oh\pa{ns}$ we can set $s = n^{1/3}$ to balance the overall number of queries.

We are ready to sketch the algorithm for partitioning a subset. We want to find a set of centers $A$ such that it partitions the set $S$ into $\tilde\Oh(t)$ Voronoi cells of sizes $\tilde\Oh(\frac{\abs{S}}t)$ for some (small enough) $t$. $W$ will be used to denote the set of vertices with large clusters (with respect to $A$). Formally, $W$ is defined as $W = \{v \in V : \abs{C_A(v)}\geq \frac{6\abs{S}}t\}$. Note that we do not require $A$ to be a subset of $S$. The following property of $W$ holds: if $A \subseteq B$, then $C_A(v) \supseteq C_B(v)$, so $W_A \supseteq W_B$.

We will be looking for $A$ with the following crucial property:
\begin{definition}
\label{def:valid}
	We will call $A$ \textnormal{valid} if $N_2(A) \cap W = \emptyset$. This means that every vertex at a distance at most $2$ from $A$ has cluster size $\Oh\pa{\frac{\abs{S}}t}$.
\end{definition}
If $A$ is \textit{valid}, the cells are small, as follows from the construction of Voronoi cells (Algorithm~\ref{alg:cells}). This is a weaker condition than the one used in~\cite{DBLP:journals/talg/KannanMZ18} and~\cite{DBLP:conf/spaa/ThorupZ01}, where the clusters of all vertices had to be small (equivalently, $W$ was required to be empty). It turns out that this relaxation enables us to use a different subprocedure. 


We will separate the search for centers into two phases, depending on the size of $W$. The parameters of used subprocedures are chosen such that the number of queries matches $\tilde\Oh\pa{\abs{S}s}$. We start with the set $\tilde W = V$, as in Algorithm~\ref{alg:basic}. If $\tilde W$ is large, a random vertex belongs there with a large probability. Thus, we can easily sample $t$ random members of $\tilde W$, add them to $A$ and halve $\tilde W$ with constant probability. Because we do not keep track of $\tilde W$, if we want to know how many vertices are in $\tilde W$ or want to sample them, then we take a sample of vertices from $V$ and check their membership of $\tilde W$. After a few iterations of the sampling, the set $\tilde W$ becomes small, and the above reasoning no longer applies: the number of queries needed to sample $t$ members of $\tilde W$ exceeds the allowed number $\tilde\Oh\pa{\abs{S}s}$. We will then end up with a set $A$ for which $\abs{W}$ is small (because $\tilde W$ is small and w.h.p. $W\subseteq \tilde W$). Then we have to switch to other procedures. Taking advantage of the above validity condition, it suffices to search for members of $W$ around the set $A$ in a BFS-like manner. That is, using BFS we can find a smallest (closure) set $W' \supseteq A$ such that any member of $W$ in the $2$-neighborhood of $W'$ is in $W'$, simply by sequentially visiting members of $W$ in the $2$-neighborhood of the visited set. This way we visit at most $\tilde\Oh\pa{\abs{W'}+\abs{A}}$ vertices. At the end, the found subset $W' \subseteq W$ is used as an input to the adapted version of the original algorithm (Algorithm~\ref{alg:thorup} from~\cite{DBLP:conf/spaa/ThorupZ01}), which turns out to be sufficient. This gives us a partition.

There is one missing component of the BFS: we have to search for neighbors effectively. The search for neighbors is naive: we iterate over cells containing the vertex and vertices in them, checking if the distance is $1$. Hence, provided that the cells are small, it suffices to ensure that the number of cells containing each vertex is also small. That is: we know how to find a valid partition but we additionally want each vertex to be in $\tilde\Oh\pa1$ cells. The general procedure (\Cref{alg:perturb}) is shown in~\Cref{sec:neighbors}: we start with a valid set of centers $A$ and fix it, returning a set $\overline{A} \supseteq A$ of roughly the same size. The set $\overline{A}$ is constructed randomly. For each vertex of $A$ with probability $\frac1{2\Delta}$ we add all its neighbors to $A$ and repeat on them, otherwise we add none of its neighbors. 
This can be seen as growing a random tree rooted at each vertex of $A$. From the point of view of an arbitrary vertex $v$, w.h.p. the endpoints of different branches of the random trees obtained from
$\overline{A}$ will not be too concentrated at the same distance from $v$. This is illustrated in the following figure, where the vertices in $A$ are orange, the vertices in $\overline{A}$ are shown in red, and other vertices from the BFS tree rooted at $v$ are shown in black. The dotted lines represent the recursion tree of the procedure while the straight lines correspond to the edges of the BFS tree.

\begin{center}
\includegraphics[scale=0.8]{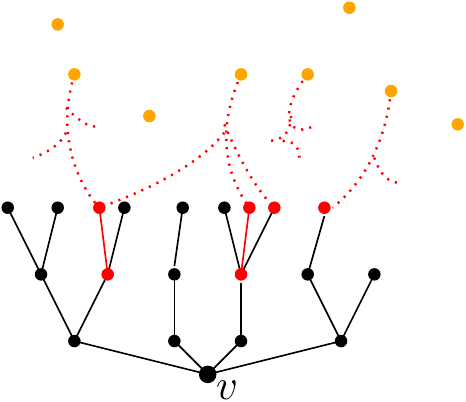}
\end{center} 

For the recursion of depth one, the spreading is performed when $W_A = \emptyset$, and then the new set $\overline{A}$ will still be valid (because $W_{\overline A} \subseteq W_A = \emptyset$).

To sum up, the version of the algorithm with recursion depth one does the following. First, at depth zero, we use the previous approach by Kannan, Mathieu and Zhou~\cite{DBLP:journals/talg/KannanMZ18}. Then we fix the cells such that each vertex is in $\tilde \Oh(1)$ cells by the method described in the previous paragraph.
Since the algorithm ensures $W = \emptyset$, this larger set of centers will be still valid. Then, we search for the edges inside each cell separately. To find a valid set $A$ for cell $S$, we first perform the procedure designed for large $\abs{\tilde W}$, which returns a set $A$ such that $W_A$ is small. After that we use
BFS to find the vertices which still have large clusters and search for the rest of $A$ among them. This finally results in a valid set of centers partitioning $S$. Note that this time we
do not ensure that each vertex is in a bounded number of depth $1$ cells, to avoid a technicality caused by the fact that this time the algorithm perturbing a set of centers can fail.

The reason for different treatment of the input sets for Algorithm~\ref{alg:perturb} (spreading the centers) is that to guarantee that this procedure works, it must receive a set of centers for which none of the vertices has a large cluster. Otherwise, it could reach that vertex, which would cause the respective cell to be large. The condition $W = \emptyset$ is indeed guaranteed after Algorithm~\ref{alg:basic} (from~\cite{DBLP:journals/talg/KannanMZ18}) but not after applying our approach for finding centers, since it relies on the relaxed condition of validity for a set of centers (\Cref{def:valid}). This creates an additional difficulty in composing the pieces together and switching from a recursion of depth $1$ to the fully recursive approach.
Informally, this is handled as follows. If the spreading fails (that is, finds a vertex with large cluster), then the found vertex is used as another source of the BFS. After returning to the BFS, we can repeat all the steps, until the perturbing is successful.


\section{Auxiliary Procedures for Finding Centers}
\label{sec:aux_procedures}
In preparation for the formal proof of \Cref{thm:main}, in this section we will establish three procedures. The goal of the first two is to find a valid set $A$: one of them is used during the phase when $W$ is large and the second when it is small. The third procedure is used to ensure that every vertex belongs to only a few cells.
In \Cref{sec:nonrecursive} we will combine these procedures to obtain a non-recursive algorithm that achieves $\tilde \Oh(n^{7/5})$ queries. Finally, in \Cref{sec:rec} we will obtain a recursive version of the algorithm that achieves $n^{4/3+o(1)}$ queries.

We recall that $s$ is a fixed threshold that remains the same throughout the whole algorithm, and will be adjusted later to optimise the overall complexity.
Our goal in this section is to design a procedure that uses, for a set $S$, only $\tilde\Oh(\abs{S}s)$ queries
(because the total size of sets at the given recursion depth is $\tilde\Oh(n)$).

Among the three randomized subroutines discussed in this section, only two are zero-error. The remaining one, namely \textsc{Estimated-Subset-Centers} (Algorithm~\ref{alg:bigW}, large $\abs{W}$) is correct with constant nonzero probability but cannot be verified easily.
However, if \textsc{Estimated-Subset-Centers} fails then we will detect this later during the BFS search.

\subsection{Small \texorpdfstring{$W$}{W}}

Take a set of centers $A$ and the set $W_A$.
We want to find a valid (see \Cref{def:valid}) superset of $A$. A very similar problem (with $A = \emptyset$) was already considered and solved by Algorithm~\ref{alg:thorup}. We claim that under appropriate assumptions, we can use the same approach. A simple case is when $W$ is small (and known). We can then simply make all possible queries between $W$ and $S$. Given these distances, the situation is similar to the one after some number of iterations of Algorithm~\ref{alg:thorup} (\textsc{Centers}). Hence, we can find, with constant probability, an appropriate superset $A_{end} \supseteq A$ for which the set $W_{end}$ is empty simply by using the same loop as in procedure \textsc{Centers}. We will show that the same approach works in a more general setting, where we do not know the set $W$. Let us state the lemma which follows from the analysis of Thorup and Zwick in~\cite{DBLP:conf/spaa/ThorupZ01}. In their proof the set $W$ can be substituted with any subset $W'$ of vertices with large clusters:

\begin{lemma}[Implicitly proven in~\cite{DBLP:conf/spaa/ThorupZ01}.]
	\label{lem:halving}
	Let $S \subseteq V$. Given a set $A \subseteq V$, let $W' \subseteq W_A$ be a set consisting of some vertices with large clusters. Let $A'$ be the set $A$ with added $t$ vertices from $W'$ chosen uniformly at random. Then, with probability at least $\frac 12$, $\abs{W_{A'} \cap W'} \leq \frac 12 \abs{W'}$.
\end{lemma}

This lemma is enough to design a new procedure. Procedure \textsc{Exact-Subset-Centers} (Algorithm~\ref{alg:smallW}) executes exactly the same loop as procedure \textsc{Centers}. Its input is a set of centers $A$ (for which there is a set of vertices with large clusters $W$) and the set $W'$. All distances between $W'$ and $S$ should be known. An important aspect of our generalization is that we do not require $W'=W$. In fact, the proof of Algorithm~\ref{alg:thorup} generalizes to the setting of Algorithm~\ref{alg:smallW}, provided that the following two properties hold:
\begin{itemize}
	\item $W'\subseteq W$ enables us to use~\Cref{lem:halving} to show that with constant probability, the size of the set $W'$ halves.
	\item $N_2(W' \cup A)\cap W \subseteq W'$ implies that the result is valid, that is $N_2(A_{end}) \cap W_{end} = \emptyset$. Indeed, let $A_{end}$ ($A \subseteq A_{end} \subseteq W' \cup A$) be the final set of centers and $W_{end}$ ($W_{end} \subseteq W$ from monotonicity) be the set $W_{A_{end}}$. Since $W_{end} \cap W'$ is empty (this is the loop termination condition), $W_{end} \subseteq W$ and $N_2(A_{end}) \subseteq N_2(W' \cup A)$, we can deduce that $N_2(W' \cup A)\cap W \subseteq W'$ implies $N_2(A_{end}) \cap W_{end} = \emptyset$, as desired.
\end{itemize}
Based on the analysis of Algorithm~\ref{alg:thorup} we deduce the following lemma:

\begin{lemma}
Assume that conditions $W' \subseteq W$ and $N_2(W' \cup A)\cap W \subseteq W'$ are satisfied and distances between $W'$ and $S$ are known. Then \textnormal{\textsc{Exact-Subset-Centers}} (Algorithm~\ref{alg:smallW}) returns a valid set of centers $A$ for $S$, such that with probability at least $\frac 14$, $A$ contains at most $\Oh\pa{t \cdot \log n}$ new members. The algorithm can be adapted into zero-error.
\end{lemma}

The last claim comes from the fact that the success of the algorithm can be verified without asking new queries: since distances are known, the sizes of cells are also known. In case of failure, the algorithm can be repeated until successful.

\begin{algorithm}
\caption{Selecting centers for a subset with given $W'$.}
\label{alg:smallW}
\begin{algorithmic}[1]
	\Function{Exact-Subset-Centers}{$S, A, W', t$}
	\LineComment{Assumes \query$(W', S)$ was already called.}
	\While{$W' \neq \emptyset$}
		\State $A' \leftarrow$ random $t$-element subset of $W'$
		\LineComment{or $A' = W'$ if $\abs{W'} < t$}
		\State $A \leftarrow A \cup A'$
		\For{$w \in W'$}
			\State $C(w) \leftarrow \{v \in S : \delta(w, v) < \delta(v, A)\}$
		\EndFor
		\State $W' \leftarrow \{w \in W' : \abs{C(w)} \geq \frac{6\abs{S}}t\}$
	\EndWhile
	\State \Return $A$
	\EndFunction
\end{algorithmic}
\end{algorithm}

Taking $W' \neq W$ will be crucial for the algorithm. It turns out that we can find an appropriate $W'$ using a small number of queries but we cannot do so with the set $W$. This can be justified as follows: the global condition defining $W$ as the subset consisting of all vertices satisfying a certain property can be relaxed to a local one. To check if $W'$ satisfies the conditions of \textsc{Exact-Subset-Centers} we only need to check the neighbors of $W'$. Intuitively, the complexity depends on the size of $W'$, not on $n$.



\subsection{Large \texorpdfstring{$W$}{W}}

\begin{algorithm}
\caption{Selecting centers for a subset by estimation. With constant probability, $\Oh( \frac{t^2n}{s\abs{S}})$ clusters will be of size $\Omega(\frac{\abs{S}}{t})$ after $\tilde \Oh(\abs{S}s)$ queries.}
\label{alg:bigW}
\begin{algorithmic}[1]
	\Function{Estimated-Subset-Centers}{$S, V, s, t$}
	\State $A \leftarrow \emptyset$
	\State $T \leftarrow K \cdot t \cdot \log n$ \Comment{$K$ is a large enough constant}
	\State $V_i \leftarrow \Call{Sample}{V, \frac{s\abs{S}}t}$, $ 1 \leq i \leq 12\log n$
	\State $\tilde W' \leftarrow \bigcup V_i$
	\For{$i = 1, 2, \ldots, 12\log n$}
		\If{$\abs{\tilde W' \cap V_i} < t$} \Comment{$|\tilde W| = \Oh\pa{\frac{t^2n}{s\abs{S}}}$}
			\State \Return $A$
		\EndIf
		\State $A' \leftarrow$ random $t$-element subset of $\tilde W' \cap V_i$ 
		\State \query$(A', S)$
		\State $A \leftarrow A \cup A'$

		\For{$v \in \tilde W'$}
			\State $X \leftarrow$ random multi-subset of $S$ of size $T$
			\State \query$(v, X)$
			\State $\tilde C(v) \leftarrow \abs{\{u \in X : \delta(u, v) < \delta(u, A)\}} \cdot \frac{\abs{S}}T$
		\EndFor
		\State $\tilde W' \leftarrow \{w \in \tilde W' : \tilde C(w) \geq \frac{5\abs{S}}t\}$
	\EndFor
	\State\Return $A$
	\EndFunction
\end{algorithmic}
\end{algorithm}

We want to design an algorithm which finds a set of centers $A$ for which the set $W$ is small (so that the algorithm from the previous subsection could be used). We will modify procedure \textsc{Estimated-Centers} (Algorithm~\ref{alg:basic}) in the following way: we will not keep track of the set $\tilde W$. Instead, procedure \textsc{Estimated-Subset-Centers} (Algorithm~\ref{alg:bigW}) in each iteration samples a large subset $V_i$ of vertices (around $\frac{s\abs{S}}t$), and checks which of them belong to $\tilde W$. Then, we sample exactly $t$ elements of $\tilde W \cap V_i$ and add them to the set of centers\footnote{This is the place in the algorithm where we cannot (as done in~\cite{DBLP:conf/spaa/ThorupZ01,DBLP:journals/talg/KannanMZ18}) sample each vertex with probability $\frac t{\abs{\tilde W}}$ because we do not know $\abs{\tilde W}$. Thus, we chose a distribution with similar properties, as discussed in~\Cref{sec:distributions}.}.
Notice that each $t$-element subset of $\tilde W$ is equiprobable, despite the two-step sampling, so it is equivalent to sampling uniformly from $\tilde W$.
If the size of $\tilde W \cap V_i$ is small (that is, $\Oh(t)$), then the size of $\tilde W$ is also likely to be small (that is, $\Oh\pa{\frac{t^2n}{s\abs{S}}}$). We then terminate, returning $A$.

There is a technicality involving checking if a vertex belongs to $\tilde W$, as we do not keep that set. The issue can be solved by sampling $V_i$ beforehand and keeping the set $\tilde W' = \bigcup{V_i} \cap \tilde W$, which is still small compared to $\tilde W$. This requires fixing the number of iterations. We cannot just put $V_i = V_1$ and keep track of $\tilde W \cap V_1$ because in the latter iterations the size of $V_i \cap \tilde W$ would be biased.

To analyse the algorithm we need the following lemma. Its proof is an adaptation of the proof of Thorup and Zwick~\cite{DBLP:conf/spaa/ThorupZ01} combined
with the proof of Kannan, Mathieu, and Zhou~\cite{DBLP:journals/talg/KannanMZ18}. Notice that the lemma gives us the guarantee about the well-defined set $W=W_A$ and the algorithm operates on a different auxiliary set $\tilde W'$. However, the change is minor because, with high probability, $W \subseteq \tilde W$. The algorithm cannot be easily converted to zero-error because the real size of $W$ or $\tilde W$ at the termination is unknown.

\begin{lemma}
Let $10\log n \leq t \leq \min\pa{s, \abs{S}}$. With probability at least $\frac15$, procedure \textnormal{\textsc{Estimated-Subset-Centers}} (Algorithm~\ref{alg:bigW}) terminates after $\tilde \Oh\pa{s \cdot \abs{S}}$ queries and outputs a set $A \subseteq V$ such that $\abs{A} = \tilde\Oh\pa{t}$ and $\abs{W_A} \leq \frac{2t^2n}{s\abs{S}}$.
\end{lemma}

\begin{proof}
Recall that we omit the subscript $A$ in $W_A$ and that $W=W_A = \{v \in V : \abs{C_A(v)}\geq \frac{6\abs{S}}t\}$. Observe that the bound on the size of $A$ follows trivially: there are logarithmically many iterations and in each of them $t$ elements are added to $A$.

Consider a coupling between the runs of procedures \textsc{Estimated-Subset-Centers} and \textsc{Estimated-Centers} (Algorithm~\ref{alg:basic}) when their random actions coincide. That is:
\begin{itemize}
	\item in each iteration they sample exactly the same $t$ vertices from $\tilde W$. If \textsc{Estimated-Subset-Centers} terminated earlier, in the coupling we assume that $t$-element samples from $\tilde W$ in the final iterations are uniformly random.
	\item the decisions whether to discard a vertex from $\tilde W'$ (or $\tilde W$, respectively) coincide (the invariant $\tilde W' = \tilde W \cap \bigcup{V_i}$ is preserved).
\end{itemize}
Recall the assumption $t \geq 10\log n$. Note that if $\abs{\tilde W} > \frac{2t^2n}{s\abs{S}}$ then $\E{\abs{V_i \cap \tilde W}} > 2t$. Hence, by Chernoff bound we have $\Prob{\abs{V_i \cap \tilde W} < t} \leq e^{-t/4} \leq \frac 1{12n\log n}$. We thus deduce that the probability that \textbf{the loop did not return $A$ until the size of $\tilde W$ became small} (at most $\frac{2t^2n}{s\abs{S}}$) is at least $1-\frac1n$.

We prove that with high probability, the number of queries is small. By a similar application of Chernoff bound, we can show that with probability $1-\frac 1n$ \textbf{each of the sets $V_i$ has size smaller than $2\frac {s\abs{S}}t$} (where $\frac {s\abs{S}}t > 10\log n$ follows from $s = n^{\Omega(1)}$ and $t \leq \abs{S}$). We can then bound the number of queries by $\Oh\pa{(12\log n)^2 \cdot T \cdot 2\cdot\frac{s\abs{S}}t} = \tilde\Oh\pa{\abs{S}s}$. The calls to \query$\pa{A', S}$ use at most $\tilde\Oh\pa{\abs{S}t}=\tilde\Oh\pa{\abs{S}s}$.

The coupling allows us to use~\Cref{lem:alg_basic_proof} for the coupled instance of \textsc{Estimated-Centers}. By union bound, with probability at least $\frac 15$ we deduce that:
\begin{itemize}
	\item the two claims in bold hold,
	\item and the properties in~\Cref{lem:alg_basic_proof} hold. In particular, $W_A \subseteq \tilde W$ after each iteration and $\tilde W = \emptyset$ after $12\log n$ iterations.
\end{itemize}
This is essentially all we need to prove the lemma. We bounded the number of queries. Moreover, emptiness of $\tilde W$ implies that \textsc{Estimated-Subset-Centers} returns a set $A$ when $\tilde W$ is small. The size of $W_A$ for this $A$ is bounded by $\abs{W_A} \leq \abs{\tilde W} \leq \frac{2t^2n}{s\abs{S}}$ (by the first claim in bold).
\end{proof}



\subsection{Finding neighbors \label{sec:neighbors}}

Throughout the algorithm, we will often need to find neighbors of a vertex. The general approach will be as follows. At each level of the recursion (or each stage of the non-recursive algorithm), the graph will be covered by some cells such that each edge belongs to at least one cell. In order to find the neighbors of a vertex, we have to iterate over the current cells containing it, and for each of them query all the vertices it contains. Denote by $T_{neighbors}$ an upper bound on the number of queries needed to find the neighbors of a vertex.
Our upper bound for $T_{neighbors}$ is the size of the largest cell times the maximum number of cells to which a vertex can belong.

\begin{algorithm}
\caption{Perturbing $A$ so that each vertex is in $\tilde\Oh(1)$ cells.}
\label{alg:perturb}
\begin{algorithmic}[1]
	\Function{Fix-Centers}{$A$}
		\State $\overline{A} \leftarrow \emptyset$
		\Function{Perturb}{$v$}
			\State $\overline{A} \leftarrow \overline{A} \cup \{v\}$
			\State $r \leftarrow \text{random number from [0,1]}$
			\If{$r \leq \frac 1{2\Delta}$}
				\For{$u \in N(v)\setminus\{v\}$}
					\State \Call{Perturb}{$u$}
				\EndFor
			\EndIf
		\EndFunction
		\For{$a \in A$}
			\State \Call{Perturb}{$a$}
		\EndFor
		\State \Return $\overline A$
	\EndFunction
\end{algorithmic}
\end{algorithm}

We want to bound the number of used queries. Up to now, we have only taken care of bounding the sizes of cells, but we have not bounded the number of cells containing a given vertex. The following procedure \textsc{Fix-Centers} (Algorithm~\ref{alg:perturb}) aims to achieve this through randomization. It receives a set of centers $A$ and spreads it slightly to a larger set $\overline A \supseteq A$. Informally, $\overline A$ will grow irregularly, and it is unlikely that many of its branches will stop exactly at the same distance from a vertex. With constant probability, each vertex lies in at most $\tilde\Oh(1)$ cells and the size of $A$ grows at most by a factor of $3$. Although $A$ is associated with the partitioned subset $S \subseteq V$, we will not use $S$ anywhere in the proof, and only refer to it when verifying the correctness.

The important thing is that in the following lemmas we \textbf{do not require $\overline A$ to be valid (see \Cref{def:valid})}. This means that it is possible that the cells are no longer small. The set of centers returned by the procedure is a superset of the input set. Hence, if $W_A$ is empty, it stays empty after calling the procedure and $S$ indeed is covered by small cells. In particular, it is the case where $A$ is the result of \textsc{Estimated-Centers} (Algorithm~\ref{alg:basic}). Therefore, we don't have to change anything in the non-recursive version of the algorithm. However, if $W_A$ is not empty, the validity of $A$ (the relation $N_2(A) \cap W_A = \emptyset$) does not imply the validity of $\overline A$ because we do not control what is the set $N_2(\overline A)$. This issue causes a technicality in the recursive algorithm discussed in~\Cref{sec:rec}.

\begin{lemma}
	The expected size of $\overline A$ is at most $2\abs{A}$.
	\label{lem:perturb_Asiz}
\end{lemma}
\begin{proof}
	Take a vertex $v$ and consider the recursion tree started by calling \textsc{Perturb($v$)}. Denote by $X_d$ the random variable that counts the number of calls at depth $d$ of the tree. By definition, $X_0 = 1$. Each call initiates in expectation $ \leq \frac 12$ calls at the next depth. Hence, $2\E{X_{d+1}} \leq \E{X_d}$. By induction, $\E{X_d} \leq \frac1{2^d}$ and $\E{X_0 + X_1 + \ldots} \leq 2\E{X_0} = 2$. The size of $\overline A$ is not larger than the total number of calls to \textsc{Perturb}, which is in expectation at most twice the number of root calls, equal to $\abs A$.
\end{proof}

Let $Q = K \cdot \Delta \cdot \log n$, where $K$ is a sufficiently large constant.

\begin{lemma}
	A vertex $u$ is contained, with probability at least $1-\frac1{n^2}$, in at most $Q^2 + Q - 2$ cells.
	\label{lem:perturb_cells}
\end{lemma}
\begin{proof}
	The case $u \in \overline A$ is trivial. From now on, assume that $u \not \in \overline A$.

	Consider any $\delta$. Consider all calls of \textsc{Perturb} at the vertices at distance $\delta$ from $u$. Let $v_1, v_2, \ldots, v_k$ be the sequence of those vertices, in the order in which \textsc{Perturb} was called on them, maybe with multiple occurrences of some vertices. We group them into consecutive blocks of size $Q$. We are interested only in full blocks of the form $v_{Qi+1}, \ldots, v_{Q(i+1)}$, where $Q(i+1) \leq k$. The probability that none of the calls in a full block recurse at least once is at most $\pa{1 - \frac{1}{2\Delta}} ^ Q \leq e^{-\frac K2 \cdot \log n} \leq n^{-4}$.
	
	With probability at least $1 - n^{-2}$, for all values of $\delta$ and for each of the first $n$ (or fewer if there are not that many) blocks, there is at least one vertex
	in the block that calls \textsc{Perturb} at least once. In particular, this happened for $\delta_{min} = \delta(u, \overline A)$. But, since there were no calls for $\delta_{min} -1$, none of the calls at distance $\delta_{min}$ recursed. Hence, the calls at depth $\delta_{min}$ do not form a full block and there are fewer than $Q$ of them. Similarly, at distance $\delta_{min} + 1$ there may be at most $Q - 1$ full blocks (since each block is related to a call on a lower layer) and at most $Q^2 - 1$ calls. Thus, with probability $1 - n^{-2}$, there are at most $Q-1$ calls at distance $\delta_{min}$ and $Q^2-1$ at $\delta_{min}+1$, $Q^2 + Q - 2$ in total. Recall that $u$ is in the cell centered at $a \in \overline A$ if and only if $\delta(a, u) \leq \delta(\overline A, u) + 1$. Since the number of such $a$ is at most the number of calls to \textsc{Perturb} at these layers, the lemma follows.
\end{proof}

\begin{lemma}
	Let $T_{neighbors}$ be the number of queries required to find neighbors of a vertex. With probability at least $\frac 14$, procedure \textnormal{\textsc{Fix-Centers}} uses $\tilde \Oh\pa{\abs{A} \cdot T_{neighbors}}$ queries and returns a set $\overline A$ such that $\abs{\overline A} \leq 3\abs A$ and each vertex $v\in S \subseteq V$ lies in at most $\Oh\pa{\Delta^2\log^2{n}}$ cells with respect to $\overline A$.
\end{lemma}

\begin{proof}
	By Lemma~\ref{lem:perturb_Asiz} and the Markov inequality, with probability at least $\frac 13$, $\abs{\overline A} \leq 3\abs A$. By Lemma~\ref{lem:perturb_cells}, with high probability each vertex lies in at most $2Q^2$ cells with respect to the set $\overline A$. Hence, with probability at least $\frac 14$, both of these conditions hold. The bound for the number of queries follows from the bound on the size of $\overline A$. 
\end{proof}

To verify the correctness, for each vertex, we can check the number of cells to which it belongs. This can be done with $\Oh\pa{\abs{\overline A} \cdot \abs{S}}$ queries. Hence, we get the following:

\begin{lemma}
	\label{lem:fix-centers}
    Procedure \textnormal{\textsc{Fix-Centers}} can be adapted into an algorithm that uses in expectation $\tilde \Oh\pa{\abs{A} \cdot \pa{T_{neighbors}+\abs{S}}}$ queries and returns a set $\overline A$ such that $\abs{\overline A} \leq 3\abs A$ and each vertex $v\in S \subseteq V$ lies in at most $\Oh\pa{\Delta^2\log^2{n}}$ cells with respect to $\overline A$.
\end{lemma}


\section{Non-Recursive Algorithm}
\label{sec:nonrecursive}

\begin{algorithm}
\caption{Non-recursive algorithm, $\tilde\Oh(ns+\frac{n^2}{s\sqrt s})$ queries.}
\label{alg:mainnonrec}
\begin{algorithmic}[1]
	\Function{Reconstruct}{$V, s$}
		\State $A \leftarrow \Call{Estimated-Centers}{V, s}$
		\State $A \leftarrow \Call{Fix-Centers}{A}$
		\State $\mathcal{D} \leftarrow \Call{Voronoi-Cells}{V, A}$
		\State $E \leftarrow \emptyset$
		\For{$S \in \mathcal{D}$}
			\State $A_S \leftarrow \Call{Subset-Centers}{S, V, s}$
			\State $\mathcal{D}_S \leftarrow \Call{Voronoi-Cells}{S, A_S}$
			\For{$X \in \mathcal{D}_S$}
				\State $E \leftarrow E \cup \Call{Exhaustive-Search}{X}$
			\EndFor
		\EndFor
		\State \Return $E$
	\EndFunction
\end{algorithmic}
\end{algorithm}

Let us present a simplified version of our main procedure. Recall first that the main procedure in~\cite{DBLP:journals/talg/KannanMZ18} first divides all vertices into cells, then performs an exhaustive search in each of them. In this section, we go one step further: we also divide the initial cells into smaller ones. In Algorithm~\ref{alg:mainnonrec} we first call \textsc{Estimated-Centers} (Algorithm~\ref{alg:basic}) to find $\tilde\Oh(s)$ Voronoi cells of sizes $\tilde \Oh(\frac ns)$. Then we perturb $A$ using procedure \textsc{Fix-Centers} (Algorithm~\ref{alg:perturb}) to ensure that each vertex lies in at most $\tilde\Oh\pa{1}$ cells. After that, for a given vertex, we can find its neighbors in $\tilde\Oh\pa{\frac ns}$ queries. We can now restrict the search for the edges to each individual cell. We will divide each initial cell into smaller cells, each of size $\tilde\Oh\pa{\frac n{s\sqrt s}}$, using Algorithm~\ref{alg:nonrec} described in the next subsection. For each of the smaller cells, we can find the edges by querying all pairwise distances.

\subsection{Non-recursive algorithm for subset}

We are now ready to show how to connect the pieces from~\Cref{sec:aux_procedures} to partition the subset $S$ into Voronoi cells. We can use at most $\tilde\Oh(n)$ queries to fit into the $\tilde\Oh(ns)$ total limit for the main procedure. Consider the following Algorithm~\ref{alg:nonrec}. The input set $S$ is a cell returned by \textsc{Estimated-Centers} (Algorithm~\ref{alg:basic}) so $\abs{S} = \tilde\Oh\pa{\frac ns}$. The algorithm returns the set of centers partitioning $S$ into cells of sizes $\tilde \Oh\pa{\frac n{s\sqrt s}}$. To ensure that the smaller cells are small enough, the algorithm picks $t = \frac{\abs{S}s\sqrt s}n$.
Indeed, we have $\frac{\abs{S}}t = \tilde\Oh\pa{\frac n{s\sqrt s}}$.

\begin{algorithm}
\caption{Finding centers when $\abs{S} = \tilde\Oh\pa{\frac ns}$ to divide $S$ into $\tilde \Oh\pa{\frac{\abs{S}s\sqrt s}n}$ cells of size $\tilde \Oh(\frac n{s\sqrt s})$.}
\label{alg:nonrec}
\begin{algorithmic}[1]
	\Function{Subset-Centers}{$S, V, s$}
	\State $t \leftarrow \frac{\abs{S}s\sqrt s}n$
	\State $A \leftarrow \Call{Estimated-Subset-Centers}{S, V, s, t}$
	\State $W' \leftarrow \emptyset$
	\For{$w \in A \cup W'$}
		\For{$v \in N_2(w)$}
			\State $\query(v, S)$ \LineComment{All pairs from $N_2(W' \cup A)\times S$ are queried}
			\State $C(v) \leftarrow \{u \in S : \delta(u, v) < \delta(u, A)\}$
			\If{$\abs{C(v)} \geq \frac{6\abs{S}}t$ and $v \not \in W'$
			}
				\State $W' \leftarrow W' \cup \{v\}$
			\EndIf
		\EndFor
	\EndFor
	\State \Return $\Call{Exact-Subset-Centers}{S, A, W', t}$
	\EndFunction
\end{algorithmic}
\end{algorithm}

First, we start with $A = \emptyset$ and $W = V$. We can run \textsc{Estimated-Subset-Centers} (Algorithm~\ref{alg:bigW}), which returns the set of centers $A$ for which $\abs{W} \leq \Oh\pa{\frac{t^2n}{s\abs{S}}} = \Oh\pa{\frac{\abs{S}s^2}{n}}$. Then we would like to plug $A$ into \textsc{Exact-Subset-Centers} (Algorithm~\ref{alg:smallW}) but to do so, we have to find a set $W'$ that satisfies the requirement $N_2(W' \cup A) \cap W \subseteq W'$. 
We can find $W'$ by local search: first we start with $W'=N_2(A) \cap W$ and then sequentially for each vertex in $W'$ we check if any of its $2$-neighbors belongs to $W$. If so, we have to add it to $W'$ and repeat. To check whether $v \in W$ we can simply call $\query(v, S)$ and recover $C(v)$. From the construction, the resulting $W'$ will satisfy the required condition and can be a parameter of \textsc{Exact-Subset-Centers}. 
We can also interpret the local search as a BFS finding the components of $N_2(A)$ in a graph formed by the  vertices in $W$ and the edges between vertices at distance at most $2$ (in $G$). Moreover, \textsc{Exact-Subset-Centers} requires that all distances between $W'$ and $S$ are known, but they were already retrieved while finding neighbors. All subroutines except \textsc{Estimated-Subset-Centers} are zero-error. Hence, the algorithm is correct with constant nonzero probability.

Now we examine the number of queries to ensure that it is at most $\tilde\Oh(n)$. For procedure \textsc{Estimated-Subset-Centers} (Algorithm~\ref{alg:bigW}) it is indeed $\tilde\Oh(\abs{S}s) \leq \tilde\Oh(n)$. The local search finds the neighbors of $(\Delta^2 +1) \cdot \pa{\abs{W'}+\abs{A}}$ vertices, therefore its complexity is $\tilde\Oh\pa{\pa{\abs{W'}+\abs{A}} \cdot \frac ns}$. As $\abs{A} = \tilde\Oh(t) = \tilde\Oh(\frac{\abs{S}s}n \cdot \sqrt{s}) \leq \tilde\Oh(\sqrt{s})$, it suffices to show that $\abs{W'}=\tilde\Oh(s)$, which is guaranteed by \textsc{Estimated-Subset-Centers}, as $\abs{W'}\leq \abs{W} = \Oh\pa{\frac{\abs{S}s^2}{n}} \leq \Oh(s)$. Procedure \textsc{Exact-Subset-Centers} (Algorithm~\ref{alg:smallW}) does not use any queries.

The above calculations were performed under the assumption that all the procedures were correct, which happens with nonzero constant probability. Therefore, we deduce the following lemma.

\begin{lemma}
    With probability at least $\frac15$, \textnormal{\textsc{Subset-Centers}} returns a valid set of centers $A$ using at most $\tilde \Oh\pa{n}$ queries. Moreover, as validity is easily tested, it can be adapted into zero-error.
\end{lemma}

\subsection{Complexity and correctness of \textnormal{\textsc{Reconstruct}} (Algorithm~\ref{alg:mainnonrec})}

Recall that \textsc{Estimated-Centers} is correct with probability $\frac 14$, and the other subroutines are zero-error. Hence, \textsc{Reconstruct} is correct with constant nonzero probability and can be adapted into a zero-error algorithm with expected complexity the same as a correct call of the original algorithm.


Each of the smaller $\tilde \Oh\pa{s\sqrt s}$ cells has size $\tilde \Oh\pa{\frac n{s\sqrt{s}}}$. Therefore, performing the exhaustive search for each of them takes $\tilde\Oh\pa{s\sqrt s \cdot \pa{\frac n{s\sqrt{s}}}^2} = \tilde\Oh\pa{\frac {n^2}{s\sqrt{s}}}$. The first two phases have complexity $\tilde\Oh\pa{ns}$ (first by~\cite{DBLP:journals/talg/KannanMZ18}, second by~\Cref{lem:fix-centers}). The total complexity is $\tilde\Oh\pa{\frac {n^2}{s\sqrt{s}} + ns}$, which for $s = n^{\frac 25}$ gives $\tilde \Oh\pa{n^{\frac 75}}$.

\section{Recursive Algorithm}
\label{sec:rec}

The non-recursive algorithm has two phases: the first partitions all vertices into cells and the second partitions the cells into smaller cells. We want to generalize this approach to a larger number of phases.

Starting from a subset equal to $V$ (and depth of recursion equal to $0$), we proceed in phases. In each phase, we are given sets $S_1, S_2, \ldots$ with the property that each edge is contained in at least one set. The purpose of the phase is to divide each of the current sets into smaller ones using Voronoi cells. That is, $S_1^1, S_1^2, \ldots \subseteq S_1$, $S_2^1, S_2^2, \ldots \subseteq S_2$, $\ldots$ are the sets given in the next phase. We want the division to be balanced: smaller cells should not be too large, and each vertex should not be contained in too many new cells. That is, when we partition a set $S$ into Voronoi cells, each element of $S$ should be contained in at most $\Oh\pa{\Delta^{2}\log^{2}n} = \tilde\Oh\pa{1}$ cells created from $S$. After some number of phases, when the size of the largest set from $S_i$ is small enough, we switch to the exhaustive search. 


The partition is implemented with~\Cref{alg:rec} (\textsc{Rec-Subset-Centers}), described in the next subsection. It turns out that after a proper choice of parameters, we can decrease the cell sizes as long as they are at least $\Omega\pa{\frac n{s^2}}$. Assume for a moment that there are no extra polylogarithmic factors in the bounds. Then, by appropriately adjusting $s$ the overall number of queries becomes $\Oh\pa{n^{4/3}}$. However, as we in fact do have such factors, we have to be more careful with the analysis. We will show that, even including the additional polylogarithmic factors, the overall number of queries is $n^{4/3+o(1)}$.

A pseudocode for the whole algorithm is given in~\Cref{sec:rec_calc}. We will prove that the number of queries used in the recursive phase at depth $d$ is $ns\cdot\pa{\Delta\log n}^{\Oh\pa {d+1}}$ in expectation and that for an appropriately chosen depth and parameter $s$, the overall complexity is $n^{4/3+o(1)}$.

\subsection{Algorithm for partitioning the subset}

We want to design an algorithm (\Cref{alg:rec}) which can be used as part of the recurrence described above, based on~\Cref{alg:nonrec} (\textsc{Subset-Centers}).

The main issue in directly generalizing the non-recursive algorithm is that it does not give us a guarantee that each vertex is in a small number of cells. Hence, as it would not search for neighbors fast, applying the previous algorithm directly would not work. What we used previously was procedure \textsc{Fix-Centers} (\Cref{alg:perturb}), which does not work correctly if we use~\Cref{alg:nonrec}.
The reason for this is that the algorithm does not give us the guarantee that $W$ is empty but that $2$-neighbors of $A$ are not in $W$, that is, $N_2(A) \cap W = \emptyset$.
However, remember that $\overline A$ is a random subset spread around $A$: it could still happen that $N_2(\overline A) \cap W \neq \emptyset$ where $\overline A$ is the result of \textsc{Fix-Centers} on $A$.

This can be fixed in the following way. First, we call subroutines as in the non-recursive algorithm and then perform \textsc{Fix-Centers}. To fix some notation, let $A$ be returned by \textsc{Estimated-Subset-Centers} (Algorithm~\ref{alg:bigW}), and let $W_A$ be the set $W$ with respect to $A$. Then, let $W' \subseteq W_A$ be found by local search, $A' \supseteq A$ be returned by \textsc{Exact-Subset-Centers} (Algorithm~\ref{alg:smallW}) called on $A$ and $W'$, and $\overline{A'} \supseteq A'$ be returned by \textsc{Fix-Centers} called on $A'$.
If $N_2(\overline{A'}) \cap W_{\overline{A'}}$ is empty, $\overline{A'}$ is a valid set of centers and we can return $\overline{A'}$. Otherwise, since $A \subseteq \overline{A'}$ and $W_A \supseteq W_{\overline{A'}}$, the intersection $N_2(\overline{A'}) \cap W_A$ is nonempty. We add all the members of $N_2(\overline{A'}) \cap W_A$ to $W'$. Then, we can return to the local search and repeat all the steps, although with the larger $W'$ instead (but still with $W'\subseteq W_A$). The local search can be started with newly added vertices only (because $A$ and $W_A$ have not changed). In
every attempt in which we have nonempty $N_2(\overline{A'}) \cap W_A$, we find at least one new member of $W_A$. Hence, we need at most $\abs{W_A}$ iterations until $\overline{A'}$ is a valid set of centers.

\begin{algorithm}
\caption{Selecting centers in the recursive algorithm. For $10\log n \leq t \leq \min\pa{s, \abs{S}}$, we divide $S$ into $\tilde \Oh(t)$ Voronoi cells of size $\tilde \Oh(\frac{\abs{S}}t)$. With constant probability, the number of queries is $\tilde \Oh(T_{neighbors} \cdot \frac{t^3n}{s\abs{S}} + \abs{S}s)$.}
\label{alg:rec}
\begin{algorithmic}[1]
	\Function{Rec-Subset-Centers}{$S, V, s, t$}
	\State $A \leftarrow \Call{Estimated-Subset-Centers}{S, V, s, t}$
	\State $W' \leftarrow \emptyset$
	\While{\textbf{true}}
		\For{$w \in W' \cup A$}\Comment{only consider new vertices}
			\For{$v \in N_2(w)$}
				\State $\query(v, S)$
				\State $C_A(v) \leftarrow \{u \in S : \delta(u, v) < \delta(u, A)\}$
				\If{$\abs{C_A(v)} \geq \frac{6\abs{S}}t$ and $v \not \in W'$}\label{line:wcheck}
					\State $W' \leftarrow W' \cup \{v\}$
				\EndIf
			\EndFor
		\EndFor
		\LineComment{All pairs in $N_2(W')\times S$ have been queried.}
		\State $A' \leftarrow \Call{Exact-Subset-Centers}{S, A, W', t}$
		\State $\overline{A'} \leftarrow \Call{Fix-Centers}{A'}$\Comment{$\tilde \Oh(\abs{A'}\cdot T_{neighbors})$}
        \If{$N_2(\overline{A'}) \cap W_{\overline{A'}} = \emptyset$}\Comment{as in line~\ref{line:wcheck}}
			\State \Return $\overline{A'}$
        \Else \LineComment{This happens at most $\abs{W_A} = \Oh(\frac{t^2n}{s\abs{S}})$ times.}
            \State $W' \leftarrow W' \cup \pa{N_2(\overline{A'}) \cap W_A}$
		\EndIf
	\EndWhile
	\EndFunction
\end{algorithmic}
\end{algorithm}

\subsection{Analysis of Algorithm~\ref{alg:rec}}

Assume that we can find the neighbors of a vertex using $\Oh(T_{neighbors})$ queries.

\textsc{Estimated-Subset-Centers} (Algorithm~\ref{alg:bigW}) uses $\tilde\Oh(\abs{S}s)$ queries and is run only once. With probability $\frac15$, it returns a set of centers $A$ for which $\abs{A} \leq \tilde\Oh\pa{t}$ and $\abs{W_A} \leq \Oh\pa{\frac{t^2n}{s\abs{S}}}$. The local search finds neighbors for the $(\Delta^2 +1) \cdot \pa{\abs{W'}+\abs{A}}$ vertices in total through all the iterations. Therefore, its complexity is $\tilde\Oh\pa{\pa{\abs{W'}+\abs{A}} \cdot T_{neighbors}} \leq \tilde\Oh\pa{\pa{\abs{W_A}+\abs{A}} \cdot T_{neighbors}} = \tilde\Oh\pa{T_{neighbors}\cdot \frac{t^2n}{s\abs{S}}}$. The algorithm uses the zero-error version of \textsc{Fix-Centers}. The calls to this procedure consist of finding neighbors for $\Oh(\abs{\overline{A'}}) = \Oh(\abs{A'}) = \tilde\Oh(t)$ vertices in expectation and last for at most $\Oh(\abs{W_A})$ iterations, so their expected complexity is $\tilde\Oh\pa{t\cdot \abs{W_A} \cdot T_{neighbors}} = \tilde\Oh\pa{T_{neighbors} \cdot \frac{t^3n}{s\abs{S}}}$. By the Markov inequality, with probability $\frac 56$ the total number of queries made by the calls to procedure \textsc{Fix-Centers} is at most $\tilde\Oh\pa{T_{neighbors} \cdot \frac{t^3n}{s\abs{S}}}$. The cost of checking emptiness of $N_2(\overline{A'}) \cap W_{\overline{A'}}$ has the same bounds as \textsc{Fix-Centers} (because its complexity depends on $\abs{\overline{A'}}$). Finally, \textsc{Exact-Subset-Centers} does not use any queries. Therefore, if both \textsc{Estimated-Subset-Centers} is correct and \textsc{Fix-Centers} uses the above number of queries (which happens with probability at least $\frac 15 \cdot \frac 56$), then  procedure \textsc{Rec-Subset-Centers} is correct, as summarized in the following lemma:

\begin{lemma}
    \label{lem:rec}
    Let $10\log n \leq t \leq \min\pa{s, \abs{S}}$. With probability at least $\frac16$, \textnormal{\textsc{Rec-Subset-Centers}} returns a valid set of centers that divides the set $S$ using at most $\tilde \Oh\pa{\abs{S} \cdot s + T_{neighbors} \cdot \frac{nt^3}{s\abs{S}}}$ queries and so that each vertex lies in $\Oh\pa{\Delta^2\log^2{n}} = \tilde\Oh\pa{1}$ cells. Moreover, as its correctness is easily tested, it can be adapted to zero-error.
\end{lemma}

\subsection{Analysis of the recurrence}

\label{sec:rec_calc}

\begin{algorithm}
\caption{Recursive algorithm with fixed depth.}
\label{alg:mainrec}
\begin{algorithmic}[1]
	\Function{Reconstruct-Rec}{$V, \dep, s$}
		\State $A \leftarrow \Call{Estimated-Centers}{V, s}$
		\State $A \leftarrow \Call{Fix-Centers}{A}$
		\State $\mathcal{C}_1 \leftarrow \Call{Voronoi-Cells}{V, A}$
		\State $R \leftarrow K \cdot \Delta^{2}\log^{2}n$ \Comment{constant from~\Cref{lem:rec}}
		\For{$d = 1, \dots, \dep$}
			\State $M_d \leftarrow \max_{S\in \mathcal{C}_d}{\abs{S}}$
			\State $t_d \leftarrow \pa{\frac{M_ds^2}{n R^d}}^{\frac13}$
			\For{$S \in \mathcal{C}_d$}
				\State $t \leftarrow t_d\cdot\frac{\abs{S}}{M_d}$ \Comment{the number of cells to divide $S$}
				\If{$t \geq 10\log n$}
					\State $A_S \leftarrow \Call{Rec-Subset-Centers}{S, V, s, t}$
					\State $\mathcal{C}_S \leftarrow \Call{Voronoi-Cells}{S, A_S}$
				\Else
					\State $\mathcal{C}_S \leftarrow \{S\}$
				\EndIf
			\EndFor
		\State $\mathcal C_{d+1} \leftarrow \bigcup_{S \in \mathcal C_d}{\mathcal C_S}$ \Comment{sets are of size $\leq R\cdot \frac {M_d}{t_d}$}
		\EndFor
		\State $E \leftarrow \emptyset$
		\For{$X \in \mathcal{C}_{\dep+1}$}
			\State $E \leftarrow E \cup \Call{Exhaustive-Search}{X}$
		\EndFor
		\State \Return $E$
	\EndFunction
\end{algorithmic}
\end{algorithm}

The main procedure in our approach is~\Cref{alg:mainrec} (\textsc{Reconstruct-Rec}). It executes the recursive division of cells in parallel, as discussed at the beginning of~\Cref{sec:rec}. The correctness of the algorithm is clear, since all used procedures are zero-error. The rest of the analysis in this section establishes $n^{4/3 + o(1)}$ complexity of~\textsc{Reconstruct-Rec} for suitably chosen $s$ and $\dep$. We also unroll the $n^{o(1)}$ factor in the complexity.

Denote $R = K \cdot \Delta^{2}\log^{2}n$, where $K$ is a constant hidden under $\Oh$ in~\Cref{lem:rec} (related to procedure \textsc{Fix-Centers}). For any vertex, after each phase, the number of cells that contain it grows by a factor at most $R$. Hence, at the depth $d$ of the recursion, each vertex is contained in at most $R^d$ cells and the total size of the sets in $\mathcal C_d$ is at most $n \cdot R^d$. Since $M_d$ is the size of the largest set in $\mathcal C_d$, the search for neighbors uses at most $T_{neighbors} = M_d \cdot R^d$ queries. We need the following two lemmas:

\begin{lemma}
	The expected number of queries at any depth $d$ of the recursion is $\tilde\Oh\pa{nsR^d}$.
	\label{lem:rec_phase}
\end{lemma}
\begin{proof}
	For $d=0$ it follows from~\Cref{lem:alg_basic_proof}. Now, let $d\geq 1$. Take any set $S \in \mathcal C_d$. By~\Cref{lem:rec} procedure \textsc{Rec-Subset-Centers} called on $S$ uses in expectation at most $\tilde \Oh\pa{\abs{S} \cdot s + T_{neighbors} \cdot \frac{nt_{S}^3}{s\abs{S}}}$ queries. We can estimate
	\begin{align*}
		T_{neighbors} \cdot \frac{nt_{S}^3}{s\abs{S}} &= M_d \cdot R^d \cdot \frac{n}{s\abs{S}} \cdot \pa*{t_{d}\cdot\frac{\abs{S}}{M_d}}^3 \\
		&= M_d \cdot R^d \cdot \frac{n}{s\abs{S}} \cdot \frac{M_ds^2}{n R^d}\cdot \frac{\abs{S}^3}{M_d^3} \\
		&= s\cdot \frac{\abs{S}^2}{M_d} \leq s \cdot \abs{S}. 
	\end{align*}
	Thus, procedure \textsc{Rec-Subset-Centers} uses $\tilde \Oh\pa{\abs{S} \cdot s}$ queries when called on $S$. Moreover, procedure \textsc{Voronoi-Cells} uses at most $\tilde\Oh\pa{\abs{S}\cdot \abs{A_S}} = \tilde\Oh\pa{\abs{S}\cdot t_S} \leq \tilde\Oh\pa{\abs{S}\cdot s^{\frac 23}} \leq \tilde\Oh\pa{\abs{S}\cdot s}$ queries. Since the sum of sizes of sets in $\mathcal C_d$ is bounded by $n\cdot R^d$, this phase takes $\tilde\Oh\pa{nsR^d}$ queries in total.
\end{proof}

\begin{lemma}
	For any $1\leq d \leq \dep$, $t_{d+1}\leq t_d^{\frac23}$, where $t_{\dep+1}$ is defined analogously.
	\label{lem:t_spada}
\end{lemma}
\begin{proof}
Take $S\in \mathcal C_d$. By the definition of $t_d$ and guarantees of Voronoi cells, the size of the largest subset in $\mathcal C_S$ obeys:
\begin{itemize}
	\item if $t_S \geq 10\log n$ it holds that $\max_{T\in\mathcal C_S}{\abs{T}} \leq 6\pa{\Delta^2+1}\cdot\frac{\abs{S}}{t_S} = 6\pa{\Delta^2+1}\cdot\frac{M_d}{t_d}\leq R\cdot\frac{M_d}{t_d}$
	\item if $t_S < 10\log n$ it holds that $\max_{T\in\mathcal C_S}{\abs{T}} = \abs{S} \leq 10\log n\frac{M_d}{t_d} \leq R\cdot\frac{M_d}{t_d}$
\end{itemize}
Thus
\[M_{d+1} \leq R\cdot\frac{M_d}{t_d}.\]
As $M_d = \frac{t_d^3nR^d}{s^2}$, we have
\[\frac{t_{d+1}^3nR^{d+1}}{s^2} \leq R\cdot\frac{\frac{t_d^3nR^d}{s^2}}{t_d}\]
or, equivalently
\[t_{d+1}^3 \leq t_d^2.\qedhere\]
\end{proof}

We can choose
$\dep = \ceil*{\frac{\log\log n}{\log\frac32}} \geq \frac{\log\log t_1}{\log\frac32}$. From~\Cref{lem:t_spada} we get $t_{\dep+1} \leq 2$
and
$M_{\dep+1} = \frac{t_{\dep+1}^3nR^{\dep+1}}{s^2} \leq \tilde\Oh\pa{\frac{n}{s^2} \cdot R^{\dep}}$.

We are ready to calculate the number of used queries. The exhaustive search uses the same queries as the procedure of finding neighbors called for every vertex (it queries a vertex with every other vertex in sets from $\mathcal C_{\dep+1}$ containing it), thus uses
\[\Oh\pa{n\cdot T_{neighbors}} = \Oh\pa{nM_{\dep+1}R^{\dep+1}} = \tilde \Oh\pa{\frac{n^2}{s^2}\cdot R^{2\dep}}\]
queries. By~\Cref{lem:rec_phase}, the number of queries used in total during the earlier phases is $\tilde\Oh\pa{nsR^{\dep}}$.
Balancing these two terms gives $s = n^\frac13\cdot R^{\frac13\cdot \dep}$ and the total complexity of the algorithm is $n^{4/3} \cdot \pa{\Delta\log n}^{\Oh\pa{\log\log n}} = n^{4/3+o(1)}$.

\bibliographystyle{alpha}
\bibliography{biblio}

\newpage
\appendix

\section{Sampling distributions}
\label{sec:distributions}

In all algorithms presented in this paper, when we want to decrease the size of $W$ or $\tilde W$, we sample $t$ of their vertices. This differs from the approaches by Thorup and Zwick~\cite{DBLP:conf/spaa/ThorupZ01} and Kannan, Mathieu, and Zhou~\cite{DBLP:journals/talg/KannanMZ18}, even in our presentation of their results. In fact, they sample each vertex from $W$ with probability $\frac t{\abs{W}}$ (or $\tilde W$, respectively) which is a distribution with very similar properties. We want to discuss why after the change of the distribution, the algorithms are still correct.

The proofs of the correctness of both procedures \textsc{Centers} and \textsc{Estimated-Centers} (\Cref{alg:thorup,alg:basic}) (\cite[Lemma~2.3]{DBLP:journals/talg/KannanMZ18}) hinge on showing that $W$ or $\tilde W$ is halved with constant probability. This in fact is proved in~\cite[Theorem~3.1]{DBLP:conf/spaa/ThorupZ01}. The same theorem is used in~\cite{DBLP:journals/talg/KannanMZ18} in their algorithm. The only place where the distribution appears in their argument is the final line of the proof of~\cite[Lemma~3.2]{DBLP:conf/spaa/ThorupZ01}. 
The property used there is that the expected value of the minimum of the chosen subset of $W = \{1, 2, \ldots, w\}$ (where $w = \abs{W}$), where we choose each element independently with probability $\frac tw$, is at most $\frac wt$.
However, this is also the case when we choose exactly $t$ numbers. The expected value can be calculated using the formula $\E X = \sum_{i=1}^{\infty}{\Prob{X \geq i}}$,
which holds for any nonnegative integer random variable $X$. There are $\binom{w}{t}$ $t$-element subsets of $\{1, 2, \ldots, w\}$, $\binom{w-i+1}{t}$ of them have minimum at least $i$, hence the expected value of the minimum can be calculated by the formula:
\[ \frac{\sum_{i=t}^{w}{\binom it}}{\binom wt} = \frac{\binom {w+1}{t+1}}{\binom wt}= \frac{w+1}{t+1} < \frac wt.\]
Hence, the slightly different distribution used in this paper has a lower expected value, which justifies its usage.
We comment that the only part of the algorithm which requires adjusting the distribution is~\Cref{alg:bigW}.

\section{Counterexample against $A \subseteq S$}
\label{sec:counterexample}

In this subsection, we provide more details on how to obtain an instance in which it is not enough to choose $A\subseteq S$. Formally, we construct a graph $G$, a subset $S \subseteq V$ and a choice of the parameter $t$ such that any $\Oh\pa{t}$ cells partitioning $S$ with centers inside $S$ contain at least one large cell. In fact, we additionally show that $S$ can be a cell arising from a partition of $G$ into Voronoi cells. In our construction a vertex $v$ lies on the boundary of $S$ and there are many vertices from $S$ close to $v$ through paths outside $S$. Then, regardless of the choice of $A$, the cell containing $v$ is large, provided that its center is close to $v$, which can be forced by the structure of the graph. An illustration of such a situation is shown in the figure below.
	
\begin{center}
    \includegraphics[scale=0.5]{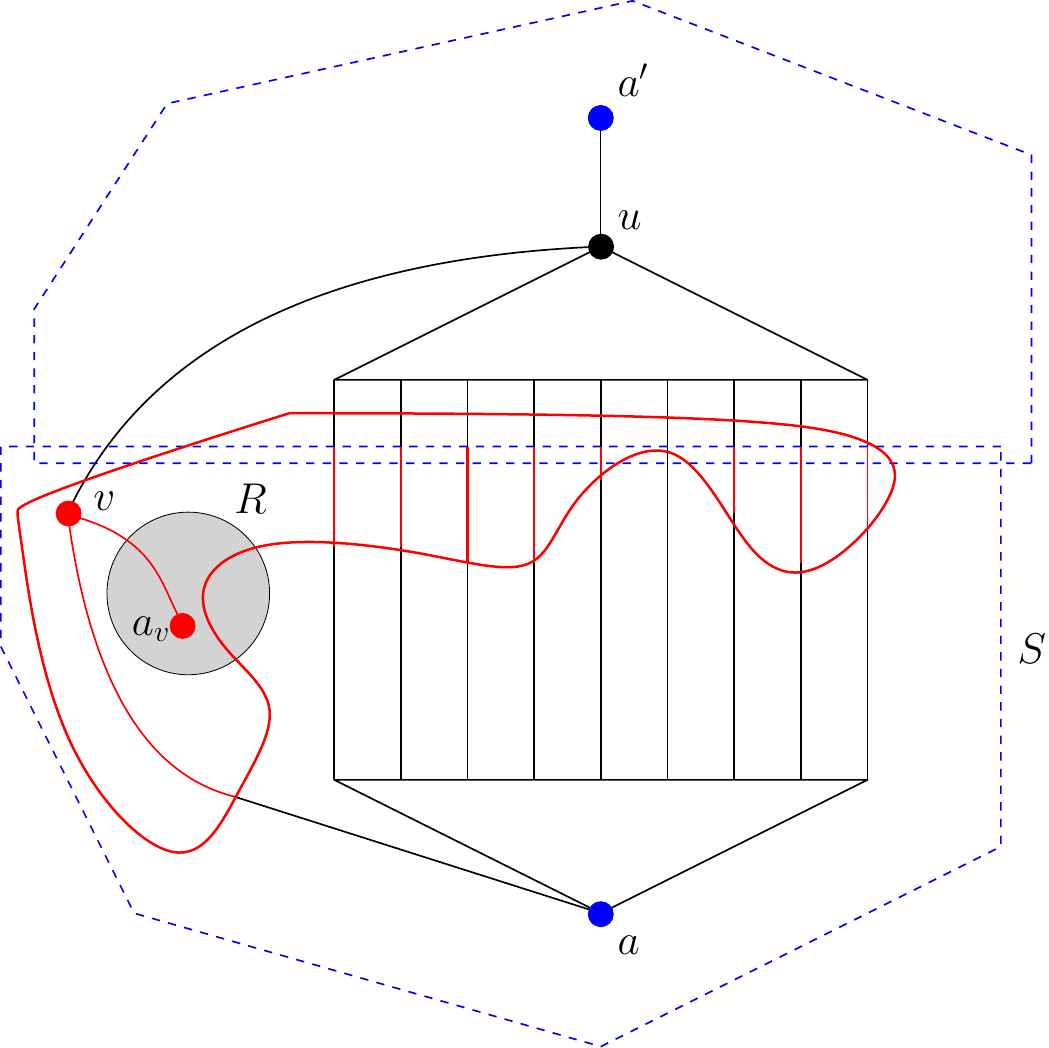}
\end{center}

In the figure, the graph $G$ is partitioned into two Voronoi cells marked in blue. Vertices $a$ and $a'$ are the centers of these cells.
The triangles represent complete binary trees, while the large segments between the two triangles denote long paths. Let the number of paths be $p$ and their length be $l$, where $\log p \ll l$.
The lines between $u$, $a'$, between $a$, $v$ and between $u$, $v$ also denote paths. By adjusting their lengths to $\frac {4l}5$, $\frac{4l}5 + \log p$ and $\log p$, respectively, we obtain that the two blue cells look as in the figure.
That is, $v$ and the bottom $\frac {9l}{10}$ vertices of all other $p$ paths are in the cell of $a$, whereas the rest of the paths and the top binary tree are in the cell of $a'$. The gray region $R$ is isolated from the rest of the graph, except for $v$.
$R$ contains many vertices and has logarithmic diameter. Denote by $S$ the cell centered at $a$. The size of $R$ can be chosen such that $\abs{R} \approx \frac{\abs{S}}2$ and $\frac{9}{10}pl \approx \frac{\abs{S}}2$. The $\approx$ sign hides $\Oh\pa{l+p}$ addends.

We claim that any partition of $S$ using at most $\frac{p}2$ \emph{Voronoi cells centered inside $S$} contains a large cell, namely the one centered at $a_v$, the chosen center that is closest to $v$.
First, we claim that $a_v$ must lie either inside $R$ or on the path between $v$ and $a$.
Indeed, the region $R$ is large so it has to be partitioned. Thus, it contains at least one center and this center has to be close to $v$ because the diameter of $R$ is logarithmic (this also implies that $\delta\pa{v, a_v} = \Oh\pa{\log p}$). Since the number of paths is $p$ and there are at most $\frac p2$ centers, not every path between triangles can contain a center.

If a path does not contain a center, then its $\frac l2 - 2\log p$ top vertices are closer to $a_v$ than to any other center. Formally, consider a path $P$ between binary trees which does not contain any center, let $P_0$ be the topmost vertex of $P$ contained in $S$ and $P_1$ the $\pa{\frac l2 - 2\log p}$-th vertex from the top. Take a center $c$ on some other path or in the lower triangle. There are two types of paths to the center $c$:
\begin{itemize}
	\item the path from $c$ through the top triangle to the vertex $P_0$ is longer than from $a_v$ to $P_0$. Indeed, the former one has length at least $\frac l5$ (because the top $\frac l{10}$ vertices of both $P$ and the path containing $c$ are outside $S$). On the other hand, the latter one has length $\frac l{10} + 2\log p$,
	\item the path from $c$ through the bottom triangle to the vertex $P_1$ is longer than from $a_v$ to $P_1$. Indeed, the former one has length at least $\pa{\frac l2 + 2\log p}$, while the latter one has length $\pa{\frac l2 - 2\log p} + \log p + \delta\pa{a_v, v} < \frac l2 + 2\log p$,
\end{itemize}
and since $a_v$ is closer than $c$ to both $P_0$ and $P_1$, it is closer to all vertices between them.
Thus, if $P$ does not contain any center, then its top $\frac l2 - 2\log p$ vertices are closer to $a_v$ than to any other center. At least $\frac l3$ of them lie in the cell of $a_v$. In total, this gives at least $\frac{lp}6 \approx \frac{\abs{S}}12$ vertices lying in the cell centered at $a_v$, which is too much. There is a cell containing a constant fraction of vertices for any $p$ with $\log p \ll l$. In fact, the result holds even for partitioning into $\Oh\pa{\frac {\abs{S}}{\log {\abs{S}}}}$ cells.

\end{document}